\documentclass[sigplan,screen]{acmart}

\copyrightyear{2019} 
\acmYear{2019} 
\setcopyright{none}
\acmConference[CPP '19]{Proceedings of the 8th ACM SIGPLAN International Conference on Certified Programs and Proofs}{January 14--15, 2019}{Cascais, Portugal}
\acmBooktitle{Proceedings of the 8th ACM SIGPLAN International Conference on Certified Programs and Proofs (CPP '19), January 14--15, 2019, Cascais, Portugal}
\startPage{1}

\usepackage{booktabs}   
\usepackage{subcaption} 

\usepackage[utf8x]{inputenc}
\usepackage{amsmath}
\usepackage{amssymb}
\usepackage{upgreek}
\usepackage{xcolor}
\usepackage{colortbl}

\usepackage{tikz}
\usetikzlibrary{arrows,positioning,calc,shapes.geometric}
\usepackage{tabularx}

\definecolor{keywordcolor}{rgb}{0.7, 0.1, 0.1}   
\definecolor{tacticcolor}{rgb}{0.1, 0.2, 0.6}    
\definecolor{commentcolor}{rgb}{0.4, 0.4, 0.4}   
\definecolor{symbolcolor}{rgb}{0.0, 0.1, 0.6}    
\definecolor{sortcolor}{rgb}{0.1, 0.5, 0.1}      

\usepackage{listings}
\usepackage{flushend}

\usepackage{amsthm}
\usepackage{underscore} 
\usepackage{array, mathtools}

\newcommand*{\carry}[1][1]{\overset{#1}}
\newcolumntype{B}[1]{r*{#1}{@{}r}}

\newtheorem*{theorem*}{Theorem}

\newcommand{\RR}{\mathbb{R}}
\newcommand{\QQ}{\mathbb{Q}}
\newcommand{\Qp}{\mathbb{Q}_p}
\newcommand{\ZZ}{\mathbb{Z}}
\newcommand{\Zp}{\mathbb{Z}_p}
\newcommand{\NN}{\mathbb{N}}
\newcommand{\CC}{\mathbb{C}}

\newcommand{\lean}[1]{\lstinline[language=lean]{#1}}

\usepackage{scalefnt} 
\begin{document}

\title[A Formal Proof of Hensel's Lemma over the \textit{p}-adic Integers]
{A Formal Proof of Hensel's Lemma over~the~\textit{p}-adic~Integers}


\author{Robert Y. Lewis}
\affiliation{
  \institution{Vrije Universiteit Amsterdam}            
  \country{The Netherlands}                    
}
\email{r.y.lewis@vu.nl}          

\begin{abstract}
The field of $p$-adic numbers $\Qp$ and the ring of $p$-adic integers $\Zp$ are essential constructions of modern number theory. Hensel's lemma, described by Gouv\^ea as the ``most important algebraic property of the $p$-adic numbers,'' shows the existence of roots of polynomials over $\Zp$ provided an initial seed point. The theorem can be proved for the $p$-adics with significantly weaker hypotheses than for general rings. We construct $\Qp$ and $\Zp$ in the Lean proof assistant, with various associated algebraic properties, and formally prove a strong form of Hensel's lemma. The proof lies at the intersection of algebraic and analytic reasoning and demonstrates how the Lean mathematical library handles such a heterogeneous topic.
\end{abstract}



\maketitle

\section{Introduction}
\label{section:introduction}

It has long been a goal of the formalized mathematics community to verify the typical undergraduate mathematics curriculum in a unified way using a single proof assistant. Various systems have achieved this goal to varying degrees \cite{Bancerek2018, harrison:hollight, mahboubi:mathcomp, megill2006, afp}, and it now seems reasonable to say that most components of this curriculum have been formalized in one system or another. Undeniably, though, some fields have received much heavier attention than others; in particular, it appears that formal developments in number theory and geometry have lagged behind those in other domains. This imbalance becomes even greater when one looks beyond undergraduate mathematics. While a few landmark projects have verified deep mathematical results \cite{gonthier2008, gonthier2013oo, hales2017}, the associated theory developments have been thin and specific to the target theorems. Research-level \emph{theorems} have been formalized, but research-level \emph{theories} remain largely untouched.

%

A recent project begun at the Vrije Universiteit Amsterdam aims to address this imbalance by bringing together traditional mathematicians, formalizers, and tool developers to work toward modern results in number theory.\footnote{\url{https://lean-forward.github.io/}} With researchers working at all levels of the theorem proving pipeline, the project will search for technology and design decisions that make it plausible to formalize deep mathematics that spans across fields. This paper breaks some initial ground to build toward this goal by constructing the $p$-adic numbers $\Qp$ and the $p$-adic integers $\Zp$ in the Lean proof assistant and verifying Hensel's lemma, a foundational result about these numbers.

The $p$-adics are a fundamental object of study in number theory with both theoretic and numeric applications. Their construction involves a mix of analytic and algebraic methods. For this reason, they make an excellent (or even necessary) point from which to embark on a project to formalize number theory. Hensel's lemma, an analogue of Newton's method for approximating roots, holds a prominent place in the study of the $p$-adics. Its computational applications make it of interest to number theorists and computer scientists alike.

In Section~\ref{section:padics}, we give an informal overview of the $p$-adic numbers and Hensel's lemma, outlining the construction and proof followed in our formalization. Section \ref{section:mathlib} briefly describes the Lean theorem prover and the mathematical library on which this project depends. Sections~\ref{section:formalpadics} and \ref{section:formalhensel} explain the formal construction of $\Qp$ and $\Zp$ and the formal proof of Hensel's lemma, respectively, focusing on design decisions made during the formalization process. In Sections~\ref{section:related} and \ref{section:conclusion}, we consider related work and reflect on the project.

The formalization described in this paper is incorporated into the Lean mathematical library, available on GitHub.\footnote{\url{https://github.com/leanprover/mathlib/}} Since this library is regularly changing, we preserve a snapshot of its status at the time this paper was submitted. This snapshot, and a map between this paper and the formalization, can be found on the author's website.\footnote{\url{http://robertylewis.com/padics/}} The code blocks presented in this paper should be read as schematic, not literal: we sometimes change names, omit universe levels, and swap implicit and explicit arguments for the sake of presentation.

\section{The \textit{p}-adic Numbers and Hensel's Lemma}
\label{section:padics}

Readers who have seen the construction of the real numbers $\RR$ via Cauchy sequences of rationals will find the construction of $\Qp$ familiar. To motivate stepping from $\QQ$ to $\RR$, we traditionally point to the fact that $\QQ$ is \emph{incomplete}: there are sequences of rational numbers that seem to approach a value but do not converge to any rational number. As an example, consider the sequence $r = (1, 1.4, 1.41, 1.414, 1.4142, \ldots)$ where the entry at index $n$ is the greatest $(n+1)$-digit rational number whose square is less than 2. This sequence has no limit in $\QQ$, since $\sqrt{2}$ is irrational. We obtain $\RR$ by creating points to represent the limit of $r$ and similarly ``convergent'' sequences.

More precisely, we say that a sequence $s : \NN \to \QQ$ is \emph{Cauchy} if for every positive $\epsilon \in \QQ$, there exists a number $N$ such that for all $k \geq N$, $|s_N - s_k| < \epsilon$. Two sequences $s$ and $t$ are \emph{equivalent}, written $s \sim t$, if for every positive $\epsilon \in \QQ$, there exists an $N$ such that for all $k \geq N$, $|s_k - t_k| < \epsilon$. Morally, a Cauchy sequence is one whose terms eventually get arbitrarily close to each other, and should converge to a (possibly irrational) value; two Cauchy sequences are equivalent if they should converge to the same value. The set of real numbers $\RR$ is defined to be the quotient of the set of Cauchy sequences with respect to $\sim$, which is to say that a real number is a set of equivalent Cauchy sequences. It can be shown that Cauchy sequences inherit the field operations from $\QQ$, and that these operations respect~$\sim$, so they can be lifted to the quotient.

We often call $\RR$ the \emph{completion} of $\QQ$ because it is the smallest extension of $\QQ$ in which all Cauchy sequences converge. But it is more accurate to say that $\RR$ is \emph{a} completion of $\QQ$, since the notion of a Cauchy sequence is parametrized by a notion of closeness. A function $f$ on $\QQ$ is a (generic) absolute value if it is positive-definite ($f(0) = 0$ and $f(k) > 0$ otherwise), subadditive ($f(x+y) \leq f(x) + f(y)$), and multiplicative ($f(x\cdot y)=f(x)\cdot f(y)$). We can replay the construction above, replacing the standard absolute value on $\QQ$ with any generic absolute value. If we use the trivial absolute value $a(x)=(0 \mathtt{\ if\ }x = 0\mathtt{\ else\ }1)$, then a sequence of rationals is Cauchy if and only if it is eventually constant, and so the completion process adds no new points to $\QQ$.

\begin{figure}
 \begin{subfigure}[t!]{0.3\columnwidth}
 \begin{equation*}
\begin{array}{B3}
    &\ldots  \mathtt{\carry 4\carry 4\carry 4\carry 4 \carry 4\carry 4\carry 4}&\mathtt{4} \\ 
      {} \hbox{$+$~} &                                                     &\mathtt{1} \\ \cline{2-3}
            &                                                    & \mathtt{0} \\
\end{array}
\end{equation*}
 \end{subfigure}
 ~
 \begin{subfigure}[t!]{0.3\columnwidth}
 \begin{equation*}
\begin{array}{B3}
    &\ldots  \mathtt{\carry 3\carry[2] 1\carry 3 \carry[2] 1\carry 3\carry[2] 1 \carry 3}&\mathtt{2} \\
      {} \hbox{$\times$~}  &                                                   &\mathtt{3} \\ \cline{2-3}
            &                                                    & \mathtt{1} \\
\end{array}
\end{equation*}
 \end{subfigure}
~
 \begin{subfigure}[t!]{0.3\columnwidth}
 \begin{equation*}
\begin{array}{B3}
    &\ldots  \mathtt{\carry 3\carry 1\carry 3 \carry 1\carry 3\carry 1 \carry 3}&\mathtt{2} \\
      {} \hbox{$+$~}  & \ldots \mathtt{4444444}                                           &\mathtt{4} \\ \cline{2-3}
            & \ldots     \mathtt{3131313}                                              & \mathtt{1} \\
\end{array}
\end{equation*}
 \end{subfigure}

 \caption{If we represent $\Qp$ as left-infinite streams of digits, we can perform addition and multiplication in base $p$ by carrying remainders to the left. $5$-adically, $\ldots 444444 + 1 = 0$ and $\ldots 313132 \times 3 = 1$. The former means that adding $\ldots 444444$ to any number is the same as subtracting $1$. All of these numbers (besides 0) have $5$-adic norm equal to 1.}
 \label{figure:padic}
\end{figure}

There are infinitely many absolute values on $\QQ$, even identifying scalar multiples: every prime number induces a unique absolute value. Fix $p \in \NN$ with $p > 1$, and for $z \in \ZZ$ with $z \neq 0$, define the \emph{$p$-adic valuation} $\nu_p(z)$ to be the largest $n$ such that $p^n \mid z$, which we read as ``$p^n$ divides $z$.'' This valuation extends to $\QQ$ by setting $\nu_p(q/r)=\nu_p(q)-\nu_p(r)$, where $q$ and $r$ are coprime. The \emph{$p$-adic norm} on $\QQ$ is defined by $|x|_p=p^{-\nu_p(x)}$ with $|0|_p=0$. If $p$ is prime, this function is an absolute value. Surprisingly, this exhausts the list of possibilities. Ostrowski's theorem \cite{ostrowski1916} states that any absolute value on $\QQ$ is (a positive scalar power of) either the standard absolute value, the trivial absolute value, or the $p$-adic norm for some prime $p$. The \emph{$p$-adic numbers} $\Qp$ are the completion of $\QQ$ with respect to the $p$-adic norm. Since a positive scalar power of an absolute value induces an isomorphic completion, we can focus our attention on $\RR$ and the family $\{\Qp \mid p \text{ prime}\}$. 

To get an intuition for how the $p$-adic numbers behave, consider what it means for two rationals to be close under the $p$-adic norm. A rational $x$ has small $p$-adic norm if $p^{-\nu_p(x)}$ is small, that is, if $\nu_p(x)$ is large, which means that $x$ is divisible by a large power of $p$. Thus $|x - y|_p$ is small if $x$ and $y$ are separated by a multiple of $p^k$ for some large $k$. The elements of a Cauchy sequence, then, are separated by multiples of larger and larger powers of $p$. 

The traditional decimal expansion allows us to write any nonzero real number in the form $\pm \Sigma_{i=-\infty}^k a_i\cdot10^i$, where $k$ is a (finite, possibly negative) integer and each $a_i \in \{0, 1, \ldots, 9\}$ with $a_k \ne 0$.
We can analogously define a $p$-adic expansion allowing us to write any nonzero element of $\Qp$ uniquely in the form $\Sigma_{i = k}^\infty a_i\cdot p^i$, with $a_i \in \{0, 1, \ldots, p-1\}$ and $a_k \neq 0$ (Figure \ref{figure:padic}). Such a series does not necessarily converge to a real number, but it does converge under the $p$-adic norm. For $m \geq n$, the difference between the $m$th and $n$th partial sums is divisible by $p^n$, so its norm is bounded by $p^{-n}$, which tends to 0 as $n$ grows.

The $p$-adic norm on $\QQ$ extends to a norm on $\Qp$, such that $|\Sigma_{i = k}^\infty a_i\cdot p^i|_p = p^{-k}$. These norms share a useful (if counterintuitive) property. The familiar triangle inequality states that $|x + y| \le |x| + |y|$ for any $x$ and $y$. But we can show something stronger for the $p$-adic norm, namely the \emph{nonarchimedean} property, which states that 
\begin{equation*}
|x + y|_p \le \max( |x|_p, |y|_p).
\end{equation*}
In fact, if $|x|_p \neq |y|_p$, it holds that $|x + y|_p = \max( |x|_p, |y|_p)$.

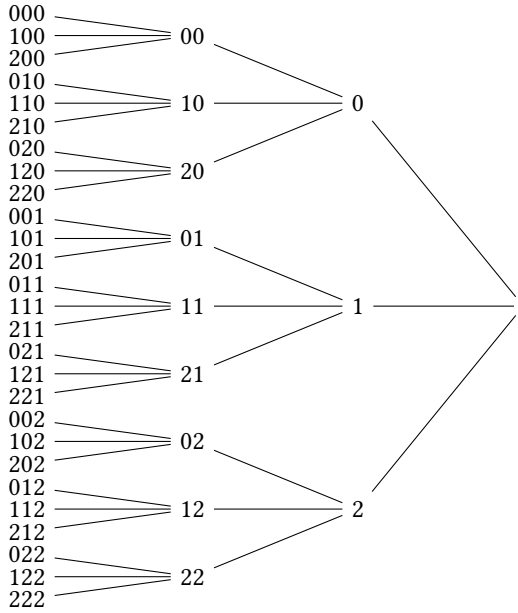
\begin{figure}
\tikzstyle{level 1}=[level distance=2.2cm, sibling distance=2.7cm]
\tikzstyle{level 2}=[level distance=2.2cm, sibling distance=.9cm]
\tikzstyle{level 3}=[level distance=2.2cm, sibling distance=.3cm]
 \begin{tikzpicture}[grow=left,node distance=13mm]
   \node{} 
    child{node{0}
       child{node{00}
         child{node{000}}
         child{node{100}}
         child{node{200}}
       }
       child{node{10}
         child{node{010}}
         child{node{110}}
         child{node{210}}
       }
       child{node(a){20}
         child{node{020}}
         child{node{120}}
         child{node{220}}
       }
     }
    child{node{1}
       child{node{01}
         child{node{001}}
         child{node{101}}
         child{node{201}}
       }
       child{node{11}
         child{node{011}}
         child{node{111}}
         child{node{211}}
       }
       child{node{21}
         child{node{021}}
         child{node{121}}
         child{node{221}}
       }
     }
    child{node{2}
       child{node{02}
         child{node{002}}
         child{node{102}}
         child{node{202}}
       }
       child{node{12}
         child{node{012}}
         child{node{112}}
         child{node{212}}
       }
       child{node{22}
         child{node{022}}
         child{node{122}}
         child{node{222}}
       }
     }
    ;
 \end{tikzpicture}   
 \caption{This diagram displays initial segments of elements of $\ZZ_3$. The $3$-adic integers are the infinite continuation of this tree to the left. The distance between two $3$-adic integers is determined by the depth of their first common ancestor, where deeper means closer. Thus, the open sets of the topology on $\ZZ_3$ are generated by the down(left)-sets of this infinite tree.}
 \label{figure:3graph}
\end{figure}

Under the standard absolute value, any Cauchy sequence of integers is eventually constant, but this is not the case under the $p$-adic norm. It can be shown that such a sequence converges to a $p$-adic number $z$ with $|z|_p \leq 1$, and so we define the \emph{$p$-adic integers} $\Zp = \{z \in \Qp \mid |z|_p \leq 1\}$ (Figure~\ref{figure:3graph}). Equivalently, a $p$-adic integer is a $p$-adic number whose $p$-adic expansion has no nonzero values to the right of the decimal point. Because of the nonarchimedean property, the sum of two $p$-adic integers is still an integer, and thus $\Zp$ forms a ring. 

As a complete structure with a nonarchimedean norm, $\Qp$ is a natural setting to develop a theory of analysis. Many of the familiar notions of calculus over $\RR$ are simplified in this setting. For instance, deciding whether an infinite series of real numbers $\sum_{i=0}^\infty r_i$ converges can be a subtle problem, and calculus students traditionally learn a list of convergence tests to answer it. Over $\Qp$, the nonarchimedean property guarantees that such a series converges if and only if $\lim_{i \to \infty} |r_i|_p = 0$. Mahler's theorem \cite{mahler1958} describes a remarkably straightforward characterization of continuous functions on $\Zp$. 

Applications of $p$-adic analysis arise in many areas of number theory, including in the studies of Diophantine equations and arithmetic progressions \cite{lech1953}. In computer science, the $p$-adics can be used to implement efficient rational arithmetic \cite{hehner1979}. The $p$-adic integers $\Zp$ are particularly useful for establishing facts about divisibility and modularity. Just as analysis over $\Qp$ is in some sense simpler than analysis over $\RR$, the algebraic structure on $\Zp$ makes these results comparatively easy to obtain. Another application is in the method of Chabauty--Coleman \cite{mccallum2012}, which can often be used to determine rational points on algebraic varieties. This method is used in the resolution of certain generalized Fermat equations \cite{dahmen2014}, closely related to the mathematics that the Lean Forward project will address.

Gouv\^ea \cite{gouvea97} cites Hensel's lemma \cite{hensel1908} as the ``most important algebraic property of the $p$-adic numbers.'' This result, which establishes a connection between the number-theoretic properties and the analysis of polynomial functions over $\Zp$, is the backbone of the study of the $p$-adics. It is often applied to prove the (non)existence of solutions to polynomial equations over various rings; in computer science, it appears in floating point rounding algorithms. Hensel's lemma is stated in the literature in many forms. The central idea is that for any univariate polynomial $f$ over $\Zp$, if one can find a point $a$ such that $f(a)$ and $f'(a)$ satisfy certain requirements, then $f$ has a unique root within a neighborhood of $a$. (We state the hypotheses explicitly in Section \ref{section:formalhensel}.)

Hensel's lemma can be used to reduce the problem of finding roots of a polynomial over $\Zp$ to the (finite) problem of finding roots over $\ZZ/p^k\ZZ$, typically for small $k$. The \emph{local-global principle}, also known as the Hasse or Hasse-Minkowski principle, is one of the central principles of Diophantine geometry \cite{serre1973}. It describes a general system that aims translating questions about roots over $\QQ$ to questions about roots over $\Qp$ and $\RR$, which are often easier to answer. A striking application of this principle shows that a quadratic form over $\QQ$ has nontrivial roots in $\QQ$ if and only if it has nontrivial roots in $\RR$ and $\Qp$ for all prime $p$. The scope and applications of the local-global principle are actively explored in number theory today; Browning \cite{browning2018} gives a survey of recent results.

\section{The Lean Mathematical Library}
\label{section:mathlib}

The Lean proof assistant, developed principally by Leonardo de Moura, was first released in 2014 \cite{demoura:14}. Lean implements a version of the calculus of inductive constructions (CIC) \cite{Coquand1990} with support for quotient types and classical reasoning. Since the release of the most recent version in 2017 \cite{lean:popl17}, there has been a concerted effort to develop mathlib, a comprehensive library for use in mathematics and computer science \cite{carneiro18}. 

Lean's mathlib is younger and smaller than similar libraries in other systems, such as Coq's Mathematical Components \cite{mahboubi:mathcomp} or Isabelle's Archive of Formal Proofs \cite{afp}, but it contains developments in many important areas of mathematics. It notably includes a proof of the law of quadratic reciprocity, a model of ZFC, and the construction of the Lebesgue measure on $\RR$.

The datatypes available in mathlib include the concrete types commonly found in mathematics, among them $\NN$, $\ZZ$, $\QQ$, $\RR$, and $\CC$; finite sets and multisets over a base type; and embeddings and isomorphisms between types. The algebraic hierarchy of mathlib is designed using \emph{type classes}, which endow a base type with extra structure in the forms of operations, properties, and notation \cite{spitters2011, wadler1989}. Lean's type class resolution mechanism automatically manages inheritance between type classes (Figure \ref{figure:typeclass}). If a type class \lstinline{T'} extends (directly or by transitivity) a type class \lean{T}, any theorem proved over \lean{T} will apply to any type that instantiates \lean{T'}. The algebraic hierarchy begins with semigroups and monoids and extends to rich structures including fields, Noetherian rings, and principal ideal domains. Van Doorn, von Raumer, and Buchholz \cite{vanDoorn2017} give a more detailed explanation of how type classes are used to define an algebraic hierarchy in Lean.

\newcommand{\nop}[1]{}
\begin{figure}

\begin{lstlisting}
class semigroup (α : Type) extends has_mul α :=
(mul_assoc : ∀ a b c : α, a * b * c = a * (b * c))

class monoid (α : Type) extends semigroup α, 
  has_one α :=
(one_mul : ∀ a : α, 1 * a = a) 
(mul_one : ∀ a : α, a * 1 = a)

class group (α : Type) extends monoid α, 
  has_inv α :=
(mul_left_inv : ∀ a : α, a⁻¹ * a = 1)

lemma one_inv (α : Type) [group α] : 
  1⁻¹ = (1 : α) :=
inv_eq_of_mul_eq_one (one_mul 1)
\end{lstlisting}

\caption{
A sample of the bottom of the algebraic hierarchy. The lemma 
\lean{one_inv} can be applied to any\!\lean{\ α} 
for which Lean can infer an instance of 
\lean{group α}.
}
\label{figure:typeclass}
\end{figure}

Topological structure is also managed using type classes. In particular, topologies on metric spaces, normed spaces, and similar structures are inherited from the topology defined on uniform spaces \cite{bourbaki1998}, of which all of these structures are instances. Topological notions such as limits and continuity are defined using filters \cite{holzl2013}, which specialize to more familiar definitions on metric or normed spaces. 

In contrast to many other libraries for CIC-based systems, mathlib does not focus on constructive mathematics. Most of the core datatypes are defined computably, making them able to be reduced in the kernel or virtual machine. But the more abstract mathematical theories freely use classical logic; these theories are mostly noncomputable. Since the system can easily track the computability of a declaration, terms that do not depend on additional axioms will still compute. 

Lean features a powerful \emph{metaprogramming} framework that allows users to write custom tactics in the language of Lean itself \cite{demoura:17}. There are a number of such tactics included in mathlib. Relevant to this project are \lean{linarith}, which proves linear inequality goals using certified Fourier-Motzkin elimination; \lean{ring}, a tactic based on Gregoire and Mahboubi's work in Coq \cite{gregoire2005} which normalizes expressions in the language of (semi)rings; and \lean{wlog}, which reduces symmetric goals to a single case. 

The development described in this paper uses a large portion of mathlib. In particular, it makes use of the concrete datatypes $\QQ$ and $\ZZ$, along with many lemmas concerning divisibility and modular arithmetic; the topology library, for properties about continuity and limits; the analysis library, for the definitions of normed rings and fields and the topological properties of these structures; the abstract algebra library, to derive additional algebraic structure on $\Zp$; and the polynomial library, which is needed even to state Hensel's lemma. This project has led to contributions to mathlib in all of these domains.

Readers unused to Lean syntax should note that explicit arguments to declarations are enclosed in parentheses \lean{()}, implicit arguments are enclosed in curly brackets \lean{\{\}}, and type class arguments are enclosed in square brackets \lean{[]}. Only explicit arguments are given by the user when applying a declaration. For instance, writing a theorem as
\begin{lstlisting}
lemma one_mul {α : Type} [group α] (a : α) : ...
\end{lstlisting}
specifies that the type \lstinline{α} is supposed to be inferred automatically (say, from the argument \lstinline{a}). The group structure on \lstinline{α}, which is introduced anonymously, should be inferred by type class resolution. In the context \lean{z : ℤ}, Lean will confirm that \lean{one_mul z} is a proof that \lean{1 * z = z}.

Another important feature of Lean syntax is its projection notation. Suppose \lean{S} is a structure (or record) type with a field \lean{val}, and \lean{t : S}. The typical way to access the \lean{val} field of \lean{t} is by \lean{S.val t}; here \lean{S.val} is a compound name, with \lean{val} living in the namespace \lean{S}. Lean also admits the abbreviation \lean{t.val}, using the period to separate a term and a name. This notation is not restricted to projections, although it is most commonly used there. In general, if a term named \lean{T.op} has been defined and \lean{t : T}, then \lean{t.op} abbreviates \lean{T.op} where \lean{t} is inserted as the first argument of type \lean{T}. For a concrete example, consider the type \lean{polynomial α} and the operator
\begin{lstlisting}
polynomial.eval : α → polynomial α → α 
\end{lstlisting}
which evaluates a univariate polynomial at an argument.
If we have \lean{F : polynomial α} and \lean{a : α}, we can use the notation \lean{F.eval a} in place of \lean{polynomial.eval a F}. This notation can be nested, e.g. to replace
\begin{lstlisting}
polynomial.eval a (polynomial.derivative F)
\end{lstlisting}
with \lean{F.derivative.eval a}.

\section{Formalizing the \textit{p}-adic Numbers}
\label{section:formalpadics}

In this section we describe the formal construction of $\Qp$ and $\Zp$ and the proofs of their associated algebraic properties. We approximately follow the presentation from Gouv\^ea~\cite{gouvea97}, although many of the ideas here are canonical in the mathematical literature. Broadly, our construction goes by the following plan:

\begin{enumerate}
 \item \label{enum:pval} Define the $p$-adic valuation on $\ZZ$, extend it to $\QQ$, and use this to define the $p$-adic norm.
 \item \label{enum:pnorm} Show that the $p$-adic norm on $\QQ$ is a non-archimedean absolute value.
 \item Define $\Qp$ as the completion of $\QQ$ with respect to the $p$-adic norm.
 \item \label{enum:qpstruct} Show that $\Qp$ inherits field operations and a norm from $\QQ$.
 \item Define $\Zp$ as a subtype of $\Qp$, and show that it instantiates various algebraic structures.
\end{enumerate}

Throughout this development, we will fix a natural number $p$. Some proofs in step \ref{enum:pnorm} assume only that $p > 1$. In the rest of the development, we work under the assumption that $p$ is prime. We manage this primality assumption using type classes, so such arguments never need to be given explicitly. In the code snippets below, we typically assume that these arguments have been fixed as parameters, and only include them in the signatures of our functions when we wish to highlight them.

The valuation and norm functions defined in step \ref{enum:pval} are total: instead of taking proofs (e.g. that $p$ is prime) as arguments, they return the value 0 when their arguments are not in the intended domain. This approach to defining partial functions is common in logics that support only total functions. Proofs of properties of these functions assume that the arguments are in the intended domain; these proofs are often inferred by the type class mechanism and are thus transparent to the user. 

\subsection{The \textit{p}-adic Valuation and Norm on \texorpdfstring{$\QQ$}{Q}}

The $p$-adic valuation $\nu_p(z)$ of an integer $z \neq 0$ is the largest $k$ such that $p^k \mid z$. This extends to $\QQ$ by setting $\nu_p(q/r) = \nu_p(q) - \nu_p(r)$ when $q$ and $r$ are coprime. We define these functions in Lean using the operator \lean{nat.find_greatest P b}, which returns the greatest \lean{n ≤ b} satisfying the predicate \lean{P}. Recall that \lean{z.nat_abs}, \lean{q.num}, and \lean{q.denom} are projection notation for \lean{int.nat_abs z}, \lean{rat.num q}, and \lean{rat.denom q} respectively.

\begin{lstlisting}
def padic_val (p : ℕ) (z : ℤ) : ℕ :=
if z = 0 then 0
else if p > 1 then 
  nat.find_greatest (λ k, (p ^ k) ∣ z) z.nat_abs
else 0

def padic_val_rat (p : ℕ) (q : ℚ) : ℤ :=
(padic_val p q.num : ℤ)-(padic_val p q.denom : ℤ)

def padic_norm (p : ℕ) (q : ℚ) : ℚ :=
if q = 0 then 0
else (p : ℚ)^(-(padic_val_rat p q))
\end{lstlisting}

These definitions are computable and can thus be evaluated on closed inputs. Note that \lean{padic_val} and \lean{padic_norm} both require the natural number \lean{p} as an explicit argument. In general, \lean{p} cannot be inferred from context. This makes it difficult to introduce generic notation for these functions, or to use $\QQ$ to instantiate type classes that depend on the norm, such as \lean{normed_field}. This complication will be resolved once we define $\Qp$, since \lean{p} will be an argument to the type of $p$-adic numbers. 

\subsection{Properties of the \textit{p}-adic Norm}

Proving the essential properties of $\nu_p$ is similarly straightforward, under the assumption that $p > 1$. The only lemmas that require $p$ to be prime are the multiplicative properties, e.g.:

\begin{lstlisting}
lemma mul {m n : ℤ} (hm : m ≠ 0) (hn : n ≠ 0) :
 padic_val p (m*n) = padic_val p m + padic_val p n
\end{lstlisting}

For the most part, the properties of \lean{padic_norm} follow from analogous properties of \lean{padic_val_rat}, which themselves follow from analogous properties of \lean{padic_val}. Lifting proofs requires some care with casts between $\NN$, $\ZZ$, and $\QQ$. 

The most involved proof in this section is the core of the later proof that the $p$-adic norm is nonarchimedean.

\begin{lstlisting}
theorem min_le_padic_val_rat_add {q r : ℚ} 
  (hq : q ≠ 0) (hr : r ≠ 0) (hqr : q + r ≠ 0) :
  min (padic_val_rat p q) (padic_val_rat p r) 
    ≤ padic_val_rat p (q + r)
\end{lstlisting}

Proving this fact requires an elementary but subtle computation. Once it is completed, the proof that \lean{padic_norm p} instantiates the \lean{is_absolute_value} type class (Figure \ref{figure:completion}) follows quickly. This instance depends on the primality of $p$, which is inferred by type class resolution.

\subsection{Completing \texorpdfstring{$\QQ$}{Q}}
\label{subsection:padics:completion}

There are many related notions of Cauchy completions in the mathematical literature, varying in the level of abstraction and in the structure on the base space. We considered a number of options for constructing $\Qp$. 

The first and most generic option was to perform the \emph{uniform completion} of $\QQ$ with respect to the uniform structure generated by the $p$-adic norm \cite{james1999}. A uniform space is an abstraction that falls somewhere in between a metric space, in which every two points are separated by a real-valued distance, and a topological space, which provides a generic but unquantified notion of ``separatedness.'' A uniform structure allows one to consider relative distances between points without assigning concrete values to these distances. Any uniform space $\alpha$ can be completed by considering the space of Cauchy filters over $\alpha$, where a Cauchy filter is a topological generalization of a Cauchy sequence. When the uniform structure on $\alpha$ is induced by a metric (or norm), this construction reduces to the completion described in Section~\ref{section:padics}. 

The uniform completion process has been formalized in Lean and could, in principle, be immediately specialized to obtain $\Qp$. However, the generality of this construction is not so amenable to a ``concrete'' number type like $\Qp$. It is quite difficult to lift operations that are not uniformly continuous, such as multiplication, from the base type to the completion space. Furthermore, one must reconcile the filter-based notion of completeness with a sequential notion in order to prove Hensel's lemma, which depends on finding a limit of a sequence of $p$-adic integers. These complications created by the generality of the uniform completion came with few upsides; it seemed more prudent to take a different approach. Regardless of the initial construction, $\Qp$ is easily instantiated as a complete uniform space after the fact.

A second option was to specialize the uniform completion to the completion of a normed structure. (We rejected the idea of using the metric completion, which falls in between, since it comes with all the disadvantages of the norm completion and more.) Under this specialization, the interface for lifting operations looks more familiar. Norm completions are not uncommon in mathematics, so although implementing an interface for this would take some initial effort, it could be reused in the future. 

\begin{figure}
\begin{lstlisting}
class is_absolute_value {α} [ordered_field α]
  {β} [ring β] (f : β → α) : Prop :=
(abv_nonneg : ∀ x, 0 ≤ f x)
(abv_eq_zero : ∀ {x}, f x = 0 ↔ x = 0)
(abv_add : ∀ x y, f (x + y) ≤ f x + f y)
(abv_mul : ∀ x y, f (x * y) = f x * f y)

parameters {α : Type} [comm_ring α]
parameters (β : Type) [ordered_field β] 
parameters (abv : α → β) [is_absolute_value abv]

def cau_seq : Type := 
{f : ℕ → α // is_cauchy abv f}

def equiv (f g : cau_seq) : Prop := 
∀ ε > 0, ∃ i, ∀ j ≥ i, abv (f j - g j) < ε

def completion : Type := quotient cau_seq equiv

instance : comm_ring completion := ...
\end{lstlisting}

\caption{A schematic depiction of the general Cauchy sequence completion. In Lean, \emph{parameters} are automatically included as arguments to declarations throughout the duration of a section.}
\label{figure:completion}
\end{figure}

Two downsides discouraged us from this approach. First, the norm referred to in an arbitrary normed space is typically real-valued. Defining $\Qp$ would thus depend on $\RR$, which is already a completion of $\QQ$ (using a different completion process). While logically sound, this approach would be pedagogically dubious, since it removes the direct analogy between $\Qp$ and $\RR$; it would also obscure the fact that the $p$-adic norm only takes rational values. A second, more practical, concern with this approach was related to the necessary type class inference. The norm on a space is generally inferred automatically, and the default instance for $\QQ$ is the traditional absolute value. 
To use the $p$-adic norm instead, we would have to locally override this instance. Doing this would be possible but could lead to complications in future developments that use the $p$-adic norm and traditional absolute value on $\QQ$ simultaneously. 

The option we elected to follow is to directly define the Cauchy sequence completion of a type $\alpha$, with respect to an absolute value on $\alpha$ taking values in an arbitrary ordered field (Figure \ref{figure:completion}). If $\alpha$ has a ring or field structure, this structure lifts immediately to the completion. This generalizes the former definition of $\RR$ in mathlib; we implement both $\RR$ and $\Qp$ as instances of this general construction. It has the added benefit that the ring operations on $\Qp$ are computable, although this property is not used in the current development. The downside to this approach is that some work must be done to connect the concrete $\epsilon$-definition of convergence to more general topological notions. This extra work can be minimized by instantiating $\Qp$ as a normed field, which we do in step \ref{enum:qpstruct}.

\begin{lstlisting}
def padic (p : ℕ) [prime p] : Type := 
completion ℚ (padic_norm p)

notation ℚ_[p] := padic p
\end{lstlisting}

The \lean{completion} function takes two explicit arguments, a type and a field-valued function on that type. An implicit argument, inferred by type class resolution, shows that the field-valued function is an absolute value. Since the $p$-adic norm is only an absolute value if $p$ is prime, it is essential to assume primality in the definition of $\Qp$. This hypothesis is also an implicit type class argument and normally will be inferred automatically.

\subsection{Operations on \texorpdfstring{$\Qp$}{Qp}}

The field operations on $\Qp$ are obtained for free through the Cauchy sequence construction. It also follows directly that $\QQ$ is embedded in $\Qp$. (Any constant sequence of rationals is Cauchy and is not equivalent to any other constant sequence of rationals, so each rational $q$ induces a unique $p$-adic number $\bar q$.) 

It takes more effort to lift the $p$-adic norm on $\QQ$ to a norm on $\Qp$. Intuitively, one might be tempted to claim that ``the norm of the limit is the limit of the norms,'' that is, that we should define the norm of a Cauchy sequence to be the limit of the norms of each entry. But since the $p$-adic norm is rational-valued, and $\QQ$ is not complete, this would unhelpfully produce a $\Qp$-valued norm. Note the contrast with the real numbers: because the linear order on $\QQ$ lifts to $\RR$, a real-valued norm makes sense. Instead, we exploit an important property of the $p$-adic norm, derived from the fact that its values lie in the set $\{0\} \cup \{p^k \mid k \in \ZZ\}$. Because these values are separated (except at 0), the norms of the entries of any (nonzero) Cauchy sequence are eventually constant.

\begin{lstlisting}
lemma stationary {f : cau_seq ℚ (padic_norm p)} 
  (hf : ¬ f ≈ 0) : ∃ N, ∀ m n, m ≥ N → n ≥ N → 
  padic_norm p (f n) = padic_norm p (f m)

def norm (f : cau_seq ℚ (padic_norm p)) : ℚ :=
if hf : f ≈ 0 then 0
else padic_norm p (f (stationary_point hf))

def padic_norm_e : ℚ_[p] → ℚ :=
quotient.lift norm norm_respects_equiv
\end{lstlisting}

Thus, the limit is indeed rational valued, and we can define a rational-valued norm on the type of Cauchy sequences. This norm respects equivalence, so it can be lifted to the quotient $\Qp$. With some work, the norm can be shown to preserve the essential properties of the norm on $\QQ$, including the nonarchimedean property. We can also check that the norm is indeed an extension of the norm on $\QQ$, meaning that for any $q \in \QQ$, $|\bar q|_p = |q|_p$.

Since we have defined $\Qp$ as the completion of $\QQ$ with respect to the $p$-adic norm, and the $p$-adic norm extends to $\Qp$, it is important to check that $\Qp$ is in fact complete with respect to its norm. We again highlight the difference here between $\Qp$ and $\RR$. Since the absolute value on $\RR$ is real-valued, as opposed to rational-valued, the arguments that $\Qp$ and $\RR$ are complete differ in significant ways. We do not use a general proof to cover both cases, since despite some structural similarity, the generalization is rather convoluted. We also prove that $\QQ$ is dense in $\Qp$---meaning that every $q \in \Qp$ is arbitrarily close to $\bar r$ for some $r \in \QQ$---similarly to how we prove the analogous statement in $\RR$, with a separate implementation to account for the different absolute value.

We have thus established that $\Qp$ is a complete field, densely embedding $\QQ$, with a nonarchimedean norm that extends the $p$-adic norm on $\QQ$. These properties uniquely characterize $\Qp$: any structure with these properties is isomorphic to $\Qp$. (We have not yet formalized this statement.) 

Finally, we instantiate $\Qp$ as a normed field. From this instance, $\Qp$ inherits a topology and uniform structure. The only complication, mentioned above, is that the generic norm of a normed ring is real-valued instead of rational-valued. But since the essential properties of the $p$-adic norm are already established, casting to $\RR$ is less troublesome here; similarly, the pedagogical concerns about using $\RR$ in the construction of $\Qp$ are no longer relevant.

From the normed field instance, we inherit the generic notation \lstinline{∥x∥} for the norm of a $p$-adic number \lean{x}. Unlike for the $p$-adic norm on $\QQ$, there is no ambiguity here about the parameter $p$, since it can be inferred from the type of \lean{x}.

\subsection{Defining \texorpdfstring{$\Zp$}{Zp}}

The $p$-adic integers $\Zp$ are traditionally defined as the subset $\{z \in \Qp \mid |z|_p \leq 1\}$. This is equivalent to the completion of $\ZZ$ using the $p$-adic norm, but for formalization purposes, the former definition is much simpler.

\begin{lstlisting}
def padic_int (p : ℕ) [prime p] : Type := 
{z : ℚ_[p] // ∥z∥ ≤ 1}

notation ℤ_[p] := padic_int p
\end{lstlisting}

The notation here is for Lean's \lean{subtype} data structure, meaning that a term \lean{z : ℤ_[p]} is a dependent pair of a term \lean{x : ℚ_[p]} with a proof that \lstinline{∥x∥ ≤ 1}. Note that this is not a ``strict'' subtype, in the sense that the term \lstinline{z} does not have type \lstinline{ℚ_[p]}; rather \lean{z.val}, the first projection of \lstinline{z}, has this type. We can move between the two types with little friction by defining a coercion from \lstinline{ℤ_[p]} to \lstinline{ℚ_[p]}. However, it is still convenient to minimize this kind of context shift, as we will discuss in Section \ref{section:formalhensel}.

From the properties of the $p$-adic norm, we obtain that $\Zp$ is closed under sums and products and show that it forms a subring of $\Qp$. This subring has algebraic structure that make it a fruitful object of study. Most fundamentally, we instantiate $\Zp$ as a normed commutative local ring with maximal ideal $\{x \in \Zp \mid |x|_p < 1\}$. We also show it is complete---meaning that any Cauchy sequence of $p$-adic integers converges to a $p$-adic integer---and that it densely embeds $\ZZ$. 

As with $\Qp$, the topology on $\Zp$ is inherited from its norm. The open sets are generated by the family of balls $\{x \in \Zp \mid |z - x|_p < \epsilon\}$, ranging over $\epsilon \in \RR_{>0}$ and $z \in \Zp$.  

\section{Formalizing Hensel's Lemma}
\label{section:formalhensel}

Hensel's lemma establishes another fundamental algebraic property of $\Zp$. This result provides simple criteria for locating $p$-adic integer roots of a polynomial; it is widely applied in $p$-adic analysis, and is also used in approximation algorithms in computer science \cite{MartinDorel2015}. The general notion of a Henselian local ring, defined to be a local ring for which Hensel's lemma holds, appears in algebraic geometry. Weaker analogues of Hensel's lemma hold over other structures, including the standard integers $\ZZ$, but the hypotheses of these analogues are harder to satisfy than those for $\Zp$.

The formal proof of Hensel's lemma follows a writeup by Conrad \cite{conrad:hensel}. Conrad's description is more concrete than Gouv\^ea's \cite{gouvea97} and avoids unnecessary detours into the group $\ZZ/p\ZZ$, although the approaches are schematically identical. We slightly modify Conrad's proof to perform as much computation as possible inside $\Zp$, without stepping into $\Qp$.

Conrad \cite[Theorem 4.1]{conrad:hensel} states Hensel's lemma as follows. Here, $\Zp[X]$ is the ring of univariate polynomials over a variable $X$ with coefficients in $\Zp$. The derivative $f'$ is the formal polynomial derivative, which does not rely on the notion of a limit.

\begin{theorem*}
 Suppose that $f(X) \in \Zp[X]$ and $a \in \Zp$ satisfy $|f(a)|_p < |f'(a)|_p^2$. There exists a unique $z \in \Zp$ such that $f(z) = 0$ and $|z - a|_p < |f'(a)|_p$. Furthermore, it holds that
 $|z - a|_p = |f(a)|_p/|f'(a)|_p$ and 
 $|f'(z)|_p = |f'(a)|_p$.
\end{theorem*}

Hensel's lemma is sometimes stated with the requirements $f(a)\equiv 0 \mod p$ and $f'(a)\nequiv 0 \mod p$. This is a weaker corollary of what we state here. The statement we have proven in Lean is a direct translation of the stronger version. 

\begin{lstlisting}
theorem hensels_lemma {p : ℕ} [hp : prime p]
  {F : polynomial ℤ_[p]} {a : ℤ_[p]} :
  ∥F.eval a∥ < ∥F.derivative.eval a∥^2 →
  ∃ z : ℤ_[p], F.eval z = 0 ∧ 
   ∥z - a∥ < ∥F.derivative.eval a∥ ∧ 
   ∥z - a∥ = ∥F.eval a∥ / ∥F.derivative.eval a∥ ∧ 
   ∥F.derivative.eval z∥ = ∥F.derivative.eval a∥ ∧ 
   ∀ z' : ℤ_[p], F.eval z' = 0 → 
     ∥z' - a∥ < ∥F.derivative.eval a∥ → z' = z
\end{lstlisting}

While Hensel's lemma can be proved in various different settings at different levels of generality, nearly all proofs follow the same approach: starting with the seed point $a$, they recursively define a sequence of approximations to the desired root, and argue that this sequence converges. This argument is typically seen as an analogy to Newton's method for finding roots of real functions.

Our proof goes by the following sketch:

\begin{enumerate}
 \item \label{enum:idents} Establish two generic polynomial identities that will be used at multiple points of the proof.
 \item Define a constant, depending only on $a$, that will be used to bound various quantities.
 \item \label{enum:recseq} Define a recursive sequence $\NN \to \Zp$, simultaneously proving bounds on the values of $f$ along this sequence.
 \item Show that this sequence is Cauchy.
 \item Show that the limit of this Cauchy sequence has the desired properties, in particular that it is a root of $f$.
 \item Show that this root is unique within a neighborhood of $a$.
\end{enumerate}

This sketch comes directly from Conrad. Our approach diverges slightly in step \ref{enum:recseq}, where we reconfigure the recursion to avoid unnecessary casts between $\Zp$ and $\Qp$. We also assume that $f(a) \neq 0$ for much of the proof, and handle this later as a (simple) degenerate case. Conrad does not make this special case explicit, but the argument fails at a crucial point if $f(a) = 0$. 

\subsection{Polynomial Identities}

Two polynomial identities are used in the proof to rewrite expressions into forms that we can more easily bound. These identities are not specific to the $p$-adic numbers.

The first identity allows us to separate components of $f(x)$ and $f'(x)$ from the expansion of $f(x + y)$.

\begin{lstlisting}
lemma binom_exp (f : polynomial α) (x y : α) :
  ∃ k : α, f.eval (x + y) = f.eval x + 
            (f.derivative.eval x) * y + k * y^2 
\end{lstlisting}

This identity follows from a similar statement on commutative semirings.

\begin{lstlisting}
def binom_exp' {α} [comm_semiring α] (x y : α) : 
  ∀ n : ℕ, ∃ k : α, 
  (x + y)^n = x^n + n*x^(n-1)*y + k * y^2
\end{lstlisting}

After inducting on $n$, this proof follows nearly automatically using Lean's \lean{ring} tactic. The only manual input is the value to instantiate \lean{k}. This value was computed using computer algebra software, and using a link to such software (e.g. \cite{lewis:17}), even this step could potentially be automated.

The second identity shows that $x - y$ divides $f(x) - f(y)$.

\begin{lstlisting}
lemma eval_sub (f : polynomial α) (x y : α) :
  ∃ z : α, f.eval x - f.eval y = z*(x - y)
\end{lstlisting}

This also follows from a similar algebraic statement, which is proved by induction and ring evaluation.

\subsection{A Bounding Value}

In the subsequent steps we will fix \lstinline{F : polynomial ℤ_[p]} and \lstinline{a : ℤ_[p]}, and assume that the inequality
\begin{lstlisting}
∥F.eval a∥ < ∥F.derivative.eval a∥^2
\end{lstlisting}
holds. (These assumptions are taken as \emph{parameters} in Lean, which are automatically inserted into declarations throughout the duration of a section.) To establish bounds on the terms of the sequence we will define in step \ref{enum:recseq}, we define an auxiliary constant \lean{T}.

\begin{lstlisting}
def T : ℝ := 
∥(F.eval a).val / ((F.derivative.eval a).val)^2∥
\end{lstlisting}

The division must take place in $\Qp$, since $\Zp$ is not a field. However, our hypothesis guarantees that \lean{T < 1}, so the quotient is in fact an integer. It is trivial to prove the following alternate characterization of \lean{T} (which uses the norm on $\Zp$), along with various simple facts about \lean{T} that will be useful for establishing bounds.

\begin{lstlisting}
lemma T_def : 
  T = ∥F.eval a∥ / ∥F.derivative.eval a∥^2
\end{lstlisting}

\subsection{Defining the Newton Sequence}

The core step of the proof of Hensel's lemma is to define a sequence $\{a_n\}$ of values that converge to the desired solution. The recursion is typically given by
\begin{align*}
 a_0 &= a \\
 a_{n+1} &= a_n - \frac{f(a_n)}{f'(a_n)}
\end{align*}
in informal texts. But without further information, this sequence lives in $\Qp$ instead of $\Zp$; we must establish properties about $f(a_n)$ and $f'(a_n)$ before concluding that $a_{n+1}$ is an integer. In an informal presentation, it is not a problem to first define the sequence in $\Qp$ and show integrality afterward. But doing so in our formal development would introduce another layer of casts, one which we would prefer to avoid. We pay a cost to avoid it: the recursion to build $\{a_n\}$ must incorporate the properties needed to prove integrality, making it slightly clumsier.

We define our induction hypothesis as follows:

\begin{lstlisting}
def ih (n : ℕ) (z : ℤ_[p]) : Prop :=
∥F.derivative.eval z∥ = ∥F.derivative.eval a∥ ∧ 
∥F.eval z∥ ≤ ∥F.derivative.eval a∥^2 * T ^ (2^n)
\end{lstlisting}

To construct our Newton sequence, we must (1) provide a value satisfying \lean{ih 0}, and (2) assuming we have a value \lean{z : ℤ_[p]} satisfying \lean{ih n z}, produce a value \lean{z' : ℤ_[p]} satisfying \lean{ih (n+1) z'}. The informal recursion indicates that our base value should be \lean{a}, and it is no trouble to prove \lean{ih 0 a}. Under the assumption \lean{ih n z}, we can check that 
\begin{lstlisting}
∥F.eval z / F.derivative.eval z∥ ≤ 1 
\end{lstlisting}
and so the recursive value 
\begin{lstlisting}
z' := z - F.eval z / F.derivative.eval z
\end{lstlisting}
is indeed an integer.

The more difficult part of this induction is to show that \lean{ih (n+1) z'} holds. While there is no deep theory needed to do this, we must calculate some chains of inequalities that, while relatively straightforward, are long and nonlinear. These computations invoke the inductive hypothesis on \lean{z}, the nonarchimedean property of the $p$-adic norm, and both polynomial identities described in step \ref{enum:idents}. It takes roughly 70 lines of Lean code to perform these computations, compared to roughly 10 in the informal presentation. Many of these computations fall under the scope of the tool Polya \cite{polya:jar} developed by the author. In the future, such a tool could be used to significantly condense this portion of our proof.

These computations are sufficient to define the following:

\begin{lstlisting}
def newton_seq (n : ℕ) : {z : ℤ_[p] // ih n z}
\end{lstlisting}

Projecting the first components, we obtain a sequence of $p$-adic integers satisfying the induction hypothesis.

\subsection{The Newton Sequence is Cauchy}

The sequence we have defined should lead us to the root of \lean{F} promised by Hensel's lemma. To reach it, we must show that \lean{newton_seq} is Cauchy, so that the completeness of $\Zp$ guarantees that a limit exists. We first establish the following lemma, which follows from another inequality computation:
\begin{lstlisting}
lemma newton_seq_dist {n k : ℕ} (hnk : n ≤ k) :
  ∥newton_seq k - newton_seq n∥ ≤ 
    ∥F.derivative.eval a∥ * T^(2^n)
\end{lstlisting}
(It is here that the special case $f(a) = 0$ diverges from the general argument.) Since \lean{0 ≤ T < 1}, Lean's analysis library makes it easy to show that the right hand side tends to 0, from which we can deduce (from general properties of sequences) that \lean{newton_seq} is Cauchy. We can thus define \lean{soln : ℤ_[p]} to be the limit of \lean{newton_seq}.

\subsection{Properties of the Limit}

From our induction hypothesis, we see that the values of \linebreak \lstinline{∥F.eval (newton_seq n)∥} tend to 0 as \lean{n} grows. It follows from the continuity of the norm (proved generally over normed spaces) that \lstinline{∥F.eval soln∥ = 0}, and thus \lean{F.eval soln = 0}, so we have  found a root. The equation
\begin{lstlisting}
∥F.derivative.eval soln∥ = ∥F.derivative.eval a∥ 
\end{lstlisting}
similarly follows from the induction hypothesis and the continuity of the norm and polynomial evaluation. A third limit argument shows that
\begin{lstlisting}
∥soln - a∥ = ∥F.eval a∥ / ∥F.derivative.eval a∥
\end{lstlisting}
which implies the (less precise but sometimes more useful) bound
\lstinline{∥soln - a∥ < ∥F.derivative.eval a∥}, the last property we sought.

The limit arguments in this section, when unfolded into the language of metric spaces, appear as frustrating manipulations of small numbers $\epsilon$ and large numbers $N$. We work to avoid as much frustration as possible by making these arguments topologically. Establishing general results about Cauchy sequences on topological spaces lets us keep the $\epsilon$--$N$ manipulations largely isolated.

\subsection{Uniqueness of the Solution}

Hensel's lemma does more than just locate a root of the polynomial $f$: it guarantees that the root is the only one within a neighborhood of the seed point $a$. The uniqueness proof follows from a short computation using the first polynomial identity from step \ref{enum:idents}.

\begin{lstlisting}
lemma soln_unique (z : ℤ_[p]) (he : F.eval z = 0) 
  (hlt : ∥z - a∥ < ∥F.derivative.eval a∥) :
  z = soln
\end{lstlisting}

When $a$ is already a root of $f$, uniqueness follows even more directly. We can thus show Hensel's lemma by a case distinction on whether $f(a) = 0$, providing $a$ as a witness in the special case and \lean{soln} in the general case.

\section{Related Work}
\label{section:related}

Although number theory is underrepresented in proof assistant libraries compared to other fields of mathematics, various projects have formalized results in this area. The following incomplete list indicates the depth and flavor of such projects.

The prime number theorem has been a popular target for formalization, verified first in Isabelle\slash HOL by Avigad, Donnelly, Gray, and Raff \cite{avigad:et:al:07} and subsequently by Harrison in HOL Light \cite{harrison2009}, Carneiro in Metamath \cite{carneiro2016}, and others. Isabelle's \emph{Archive of Formal Proofs} contains a number of related entries, including Eberl's proof of Dirichlet's theorem \cite{eberl:dirichlet:afp}. Elliptic curves and their number theoretic consequences have been addressed in multiple formalizations, including by Bartzia and Strub~\cite{bartzia2014}. The transcendence of $e$ and $\pi$, a result that is at least adjacent to number theory, was first formalized in Coq by Bernard, Bertot, Rideau, and Strub~\cite{Bernard:2016}. Analyses of the solutions to Pell's equations have been formalized in various systems, as has the proof of Fermat's little theorem; these and other classical results appear on Wiedijk's list of formalizations of 100 fundamental theorems \cite{wiedijk:100}.

The only formal construction of $\Qp$ and $\Zp$ found in the literature is by Pelayo, Voevodsky, and Warren \cite{pelayo15}, carried out in the Coq UniMath library \cite{UniMath}.  Because of the univalent foundations of UniMath, it is difficult to compare their approach with ours. One immediate difference is that Pelayo et al.\ begin with an algebraic construction of $\Zp$ rather than an analytic construction of $\Qp$. This construction defines $\Zp$ as a quotient on the ring of formal power series of $\ZZ$, and goes on to define $\Qp$ as the field of fractions on $\Zp$. The algebraic approach is perhaps more appropriate in a univalent setting. A complete theory of the $p$-adics ultimately requires both analytic and algebraic structure. No matter which is chosen for the initial construction, the properties of the other must eventually be derived. The UniMath development ends soon after the ring and field structures are defined, and does not prove any theorems about $\Zp$ or $\Qp$.

An undocumented construction of $\Qp$ by Harrison is also found in the HOL Light repository \cite{harrison:hollight}, where it is written that the development is ``meant as an example of using metric space completion.'' The development defines $\Qp$ as the metric completion of $\QQ$, which we believe is better avoided for pedagogical reasons (Section \ref{subsection:padics:completion}). Since metric space completion does not preserve the field operations on $\QQ$, much of the construction is dedicated to redefining these operations on $\Qp$. This development ends once the field structure on $\Qp$ is established, and does not prove any results about the type. It is interesting to note how the construction in a simply typed logic differs from those in dependent type theory. HOL Light does not allow one to define the type $\Qp$ depending on $p$ (nor on a proof that $p$ is prime). Instead, Harrison defines a general type \lean{padic} that contains the image of $\Qp$ for each~$p$. Such an approach is common in HOL-based systems, but is rarely used in systems like Lean, where the dependencies pose no problems.

Martin-Dorel, Hanrot, Mayero, and Th\'ery describe a Coq formalization of Hensel's lemma for the standard integers, and show some of its applications in verifying bounds on rounding errors \cite{MartinDorel2015}. Their approach is more explicitly algorithmic than ours, as their applications involve computing solutions to polynomials modulo powers of $p$ (a process known as Hensel lifting). For this purpose, it is reasonable to operate over the standard integers, since there are complications in defining $\Qp$ as a computable field. The statement of Hensel's lemma is rather less elegant than over $\Zp$, however, and its import as an algebraic property is hidden. 

\section{Concluding Thoughts}
\label{section:conclusion}

The $p$-adic numbers and Hensel's lemma are an important tool of modern number theory, including the study of Diophantine equations. A recent project begun at the Vrije Universiteit Amsterdam aims to formalize results in this area. Constructing $\Qp$ and $\Zp$ are essential steps toward this goal. We plan to pursue consequences of Hensel's lemma in future work, beginning with applications of the local-global principle. The $p$-adic numbers can be abstracted in different ways, and Hensel's lemma proved in more general contexts; we plan to explore these possibilities. More concretely, it is often useful to consider the alternative characterization of $\Zp$ as the inverse limit of the rings $\ZZ / p^n\ZZ$. We plan to show the connection between this algebraic approach and our analytic approach.

This development has also served as a case study for using Lean for such a project. The mix of analysis, algebra, and concrete computation make the $p$-adic numbers an interesting target; we found that Lean and its libraries were up to the task. Similar developments could certainly be made using other proof assistants, but we found the classical library and notational features of Lean to be quite helpful here; dependent type theory is a convenient logical foundation to use, since $\Qp$ is naturally a dependent type. Two particularly painful parts of the project were managing casts between various number structures and proving long but straightforward nonlinear inequalities. The former is especially likely to occur in further number theoretic developments, since it is often necessary to move between $\NN$, $\ZZ$, $\QQ$, $\RR$, $\Qp$, $\Zp$, and other number structures. We hope to develop tools to assist with this movement using Lean's metaprogramming capabilities.

As always, it is difficult to directly compare the lengths of our formalization and the informal proofs we followed, since they begin at different levels of background knowledge. The portion of our formalization that proves Hensel's lemma is around 400 lines of code, corresponding to 1.5 pages of text in Conrad's informal proof; tools mentioned in the previous paragraph should significantly decrease this ratio. This development has added around 4500 lines of code to the mathlib repository.

\begin{acks}                            
  We would like to thank Jeremy Avigad, Alexander Bentkamp, Jasmin Blanchette, Kevin Buzzard, Sander Dahmen, Johannes H\"olzl, Petar Vukmirovi\'c, and the Lean mathlib community for their support, advice, and comments. We acknowledge support from the European Research Council (ERC) under the European Union's Horizon 2020 research and innovation program (grant agreement No. 713999, Matryoshka).
\end{acks}

\bibliography{citations}

%

\end{document}